\documentstyle[prl,aps,multicol]{revtex}
\newcommand{\ea}{{\it et al.}}

\begin{document}
\draft
\title{Ion-trap quantum logic using long-wavelength radiation} 
\author{Florian Mintert$^1$ and Christof Wunderlich$^{2*}$}
\address{
$^1$ I. Institut f\"ur Theoretische Physik, Universit\"at Hamburg, Jungiusstr.9, 20355 Hamburg, Germany\\
$^2$ Institut f\"ur Laser-Physik, Universit\"at Hamburg, Jungiusstr.9, 20355 Hamburg, Germany}

\date{June 25, 2001}
\maketitle

\begin{abstract}
A quantum information processor is proposed that combines experimental 
techniques and technology successfully demonstrated either in 
nuclear magnetic resonance experiments or with  trapped ions. 
An additional inhomogenenous magnetic field applied to an ion trap 
 i) shifts individual ionic resonances (qubits), 
making them distinguishable by frequency,  and, ii) mediates the 
coupling between internal and external degrees of freedom of trapped ions.
This scheme permits one to individually 
address and coherently manipulate  
ions confined in an electrodynamic trap using radiation in the 
radiofrequency or microwave regime.
\end{abstract}


\begin{multicols}{2}

Quantum information processing (QIP) holds the promise of 
extending today's computing capabilities to problems that,
with increasing complexity, require exponentially growing resources 
in time and/or the number of physical elements \cite{gruska}.
The computation of properties of quantum systems themselves is particularly
suited to be performed on a quantum computer, even on a device where logic
operations can only be carried out with limited
precision\cite{risq}.
Elements of quantum logic operations have been successfully demonstrated
in experiments using ion traps\cite{sackett,naegerl1,huesmann},
cavity quantum electrodynamics \cite{maitre}
and in the case of nuclear magnetic resonance (NMR) even algorithms
have been performed\cite{chuang}.
Whereas quantum computation with nuclear spins in macroscopic ensembles
can most likely not be extended beyond about 10 qubits 
(quantum mechanical two-state systems) \cite{jones2},
ion traps do not suffer from limited scalability
{\it in principle} and represent a promising system to explore
QIP experimentally. 
They can be employed to also investigate fundamental questions
of quantum physics, for example related to decoherence\cite{brune}
or multiparticle entanglement\cite{sackett}.
However, they still pose considerable experimental 
challenges.

Two internal states of an {\it individual} ion are used as a qubit.
The vibrational motion of a {\it collection} of trapped ions 
serves as the ``bus-qubit" and permits conditional dynamics between 
individual qubits \cite{Cirac95}.  In order to couple internal and motional 
degrees of freedom of a trapped atom, the atom has to experience 
an appreciable variation of the field that drives the
internal transition over the extent of its spatial wavefunction.
A measure for the strength 
of the field gradient relative to the atoms
spatial extend  \mbox{$\Delta z\equiv\sqrt{\frac{\hbar}{2m\omega_l}}$}
is the Lamb-Dicke  parameter (LDP) $\eta = \Delta z \, 2\pi/\lambda$.
($\lambda$ is the wavelength of the applied radiation; 
the atom with mass $m$ is trapped
in a harmonic potential characterized by angular frequency $\omega_l$.)
For typical qubit transitions and useful trap frequencies, this parameter has an  
appreciable nonzero value only for driving radiation in the optical domain. 

Consequently, all schemes for ion trap QIP used
(for example, \cite{sackett,naegerl1,huesmann}) 
and
suggested\cite{cirac,steane}  have in common that laser light is necessary
to drive qubit transitions. 
Involved optical setups are required to cool the vibrational motion of the ions, 
and to prepare, coherently
manipulate, and readout the qubit states. 
It is desirable to find simpler methods for the manipulation of  
well isolated qubits in ion traps, methods that require 
a smaller number of laser beams and sources, and
are less demanding regarding the specifications of beam quality and 
pointing stability, and frequency and intensity stability.
 
Another important issue when trying to implement a quantum information processor
 and prerequisite for further studies using several ions is the addressing
of individual qubits out of a large collection of ions.
In order to perform operations on an optically driven transition between
qubit states of an individual ion, strongly focused laser light must
be aimed at only the desired ion\cite{naegerl2}.
Different approaches have been used and proposed instead 
\cite{turchette}
to circumvent
practical and fundamental difficulties arising from such an addressing scheme.

Techniques for generating radiation with long coherence time 
that are experimentally challenging and/or require intricate setups 
in the optical domain are well established in the radiofrequency (rf) 
or microwave (mw) domain
where commercial off-the-shelf components can be used.
Technological resources
developed over decades in this frequency range have been used 
in an inventive way for NMR methods and contributed to 
the impressive and fast success of  NMR in QIP. 
It would be desirable to use these resources
for ion trap QIP, too.
Two obstacles have precluded rf or 
mw radiation from being used for the manipulation of 
qubits in ion traps (for example, comprised of 
two hyperfine states): i) the LDP is essentially zero for 
mw radiation and useful trap frequencies. Thus, coupling
of internal and external degrees of freedom is not possible. 
ii) mw radiation cannot be focused such that
individual ions can be addressed.

Here, we show that an additional magnetic field gradient applied to an
electrodynamic trap (i) introduces a coupling between internal
and motional states even for rf or mw radiation and (ii)
serves to individually shift ionic qubit resonances thus
making them distinguishable in frequency space.  With the
introduction of this field, all optical schemes devised for
QIP in ion traps can be applied in the rf or mw regime, too.

We consider ionized atoms confined in the initially field free region
along the symmetry axis of an ac-quadrupole field of a linear electrodynamic
 trap\cite{ghosh}.
The ions are trapped due to a pseudopotential,
that is harmonic in the center of the trap and described by
$V=\frac{1}{2}m\omega_r^2(x^2+y^2)+\frac{1}{2}m\omega_z^2z^2$,
where 
the angular frequencies 
$\omega_r$ and $\omega_z$ characterize the trapping potential
in radial and axial direction, respectively.
If more than one ion is trapped, the equilibrium positions are determined by
the condition that trapping force and Coulomb forces add to zero for each ion.
As long as
$\omega_r / \omega_z \gtrsim 0.73\,N^{0.86}$,
where $N$ is the number of ions in the trap, the $x$- and $y$-components
of the equilibrium positions vanish\cite{steane} and the ions form
a linear chain characterized by axial vibrational eigenfrequencies
$\omega_l$ $(l=1,...,N)$.
Such a linear configuration will be considered in what follows.

Applying a magnetic field $\vec{B}_{dc}$ to the trap leads to Zeeman energies
$\varepsilon_0(B_{dc})$ and $\varepsilon_1(B_{dc})$
of  the internal qubit states $|0\rangle$ and $|1\rangle$.
$\vec{B}_{dc}=(bz+b_0)\hat{z}$ is chosen with magnetic field gradient
$b\equiv\frac{\partial B_{dc}(z)}{\partial z}\neq 0$ and constant offset
$b_0$, leading to an individual, position dependent Zeeman shift
for each ion such that the qubit resonance frequency
$\omega(z)=\left[\varepsilon_1(B_{dc}(z))-\varepsilon_0(B_{dc}(z))\right]\hbar^{-1}$.
With $\partial_z \varepsilon_1 \neq \partial_z \varepsilon_0$, 
this Zeeman shift gives rise to a state dependent force in a 
inhomogeneous magnetic field. 
Thus internal state transitions
cause a slight displacement of the ion and internal and motional degrees of freedom 
are coupled.
Since the spatial excursion 
of an ion is of the order $\Delta z \sqrt{2\bar{n}+1}$ 
( $\bar{n}$ is the mean vibrational quantum number at the Doppler limit),
this additional Zeeman potential is linear in the ion's displacement 
to very good approximation.  

As in the proposal by Cirac and Zoller\cite{cirac},
a collective vibrational mode is employed as means of communication (bus-qubit)
between otherwise isolated internal qubit states of the ions.
The Hamiltonian
$H=\frac{1}{2}\hbar\omega(z)\sigma_z+\hbar\omega_la_l^\dagger a_l$
describes the qubit states of a particular ion coupled to vibrational mode $l$
($\sigma_z=|1\rangle\langle1|-|0\rangle\langle0|$.)
When expanded to first order in the axial position operator,
$\zeta\Delta z (a_l^\dagger +a_l)$ it reads
\begin{equation}
H=\frac{1}{2}\hbar\omega_0\sigma_z+\hbar\omega_la_l^\dagger a_l+
\frac{1}{2}\hbar\omega_{l}\varepsilon_c(a_l^\dagger +a_l)\sigma_z
\end{equation}
with
$ 
\varepsilon_c\equiv\zeta\frac{\Delta z|\partial_z\varepsilon_1-\partial_z\varepsilon_0|}{\hbar\omega_l}.
$
Here,   
$\zeta$ is the expansion coefficient of the displacement of the ion
to be addressed in terms of the normal mode coordinate.
For the center-of-mass mode $\zeta=\frac{1}{\sqrt{N}}$
and for any other mode $\zeta\approx\frac{1}{\sqrt{N}}$.
The qubit's resonance frequency at its equlibrium position is denoted
by $\omega_0$. 
It is useful to perform the unitary transformation
$\tilde{H}=e^SHe^{-S}$ with
$S=\frac{1}{2}\varepsilon_c(a_l^\dagger-a_l)\sigma_z$
and after dropping constant terms we obtain
$
\tilde{H}=\frac{1}{2}\hbar\omega_0\sigma_z+\hbar\omega_la_l^\dagger a_l,
$
{\it i.e.} in the transformed Hamiltonian the coupling between
internal degree of freedom and vibrational mode has been eliminated.
The transformed operators are given by
\mbox{$\tilde{a}_l          =  a_l-\frac{1}{2}\varepsilon_c\sigma_z$},
\mbox{$\tilde{a}_l^\dagger  =  a_l^\dagger-\frac{1}{2}\varepsilon_c\sigma_z$},
\mbox{$\tilde{\sigma}_+    =  \sigma_+e^{ \varepsilon_c(a_l^\dagger-a_l)}$},
and
\mbox{$\tilde{\sigma}_-     =  \sigma_-e^{-\varepsilon_c(a_l^\dagger-a_l)}$.}

When an ion interacts with an additional electromagnetic field of frequency
$\omega_M$, this leads to an interaction term
\begin{equation}
H_M=\frac{1}{2}\hbar\Omega_R(\sigma_++\sigma_-)
\left(e^{i(\eta(a_l+a_l^\dagger)-\omega_Mt)}+ \mbox{h.c.}
\right),
\end{equation}
where $\eta=\zeta\sqrt{\frac{\hbar k_z^2}{2m\omega_l}}$
is the LDP for $N$ ions and
$\Omega_R=\frac{\vec{\mu}\cdot\vec{B}_M}{\hbar}$
is the Rabi frequency (here of a magnetic dipole transition)
characterizing the coupling strength.
The magnetic dipole moment operator is denoted by $\vec{\mu}$,
the magnitude of the wave vector in axial direction
$k_z=\frac{\omega_M}{c}\cos\theta$,
where $\theta$ is the angle between the incident beam and the trap axis,
and $B_M$ is the magnetic amplitude of the electromagnetic field.

The transformed interaction  $\tilde{H}_M=e^SH_Me^{-S}$ is given by
\end{multicols}
\begin{equation}
\tilde{H}_M=\frac{1}{2}\hbar\Omega_R
\left(\sigma_+e^{\varepsilon_c(a_l^\dagger-a_l)}+
\sigma_-e^{-\varepsilon_c(a_l^\dagger-a_l)}\right)
\left(e^{i(\eta(a_l+a_l^\dagger-\varepsilon_c\sigma_z)-\omega_Mt)}+
e^{-i(\eta(a_l+a_l^\dagger-\varepsilon_c\sigma_z)-\omega_Mt)}\right).
\end{equation}
\begin{multicols}{2}
It is useful to perform a further transformation to the interaction picture
with respect to $\tilde{H}$.
With detuning $\Delta=\omega_M-\omega_0$, this leads to
\end{multicols}
\begin{equation}
\tilde{H}_M=\frac{1}{2}\hbar\Omega_R
\left(\sigma_+e^{-i(\Delta t+2\eta\varepsilon_c)}
e^{i((\eta+i\varepsilon_c)a_l+(\eta-i\varepsilon_c)a_l^\dagger)}+\mbox{h.c.}\right),
\end{equation}
\begin{multicols}{2}
where terms oscillating with frequency $\pm(\omega_M+\omega_0)$
have been dropped (rotating wave approximation).
For $\varepsilon_c>0$, {\it i.e.} when a magnetic field gradient is applied,
the LDP $\eta$ can be replaced by a complex one,
$\eta+i\varepsilon_c$.
This complex parameter can be decomposed into its absolute value
$\eta'=\sqrt{\eta^2+\varepsilon_c^2}$ and its phase,
that in turn can be accounted for by incorporating it into the arbitrary
initial conditions of the phonon operator's time dependence.
Because $\sigma_+$, too is defined only up to an arbitrary phase,
the phase factor $e^{-2i\eta\varepsilon_c}$ can be appended to
this operator and what remains is the usual field-ion interaction
governed by an {\em effective} LDP $\eta'$.

When mw
radiation is used to drive internal transitions of a qubit 
in a {\em usual} ion trap 
(i.e. without magnetic field gradient), then the LDP, 
$\eta$ is very small ($\eta\approx 7 \times 10^{-7}$ for 
40 Yb$^+$ ions with
transition frequency $\omega_0 = 2\pi$ 12.6 GHz
at a trap frequency of $2\pi$ 100 kHz). Thus, 
coupling internal and external degrees of freedom of an ion
is not possible with mw radiation in the usual scheme. 
However, it {\it is} possible with an additional magnetic
field gradient: Even when $ \eta \approx 0$, then still
$\eta' \approx \varepsilon_c > 0$. 
All operations (including, for example, sideband cooling) 
that require coupling between internal states and vibration
of the ion string, usually carried out with optical fields, 
can now be implemented using microwave radiation.
In table 1 some values of $\varepsilon_c$ are listed.
The required values for $|\partial_z\varepsilon_1-\partial_z\varepsilon_0|$
will be considered in what follows.

In addition to coupling internal and external degrees of freedom of the ions,
the field gradient applied to the ion trap serves to distinguish qubits
by separating their resonance frequencies.
The magnitude of the magnetic field gradient determines
the frequency separation of qubit-resonances in adjacent ions:
The resonance frequency of a particular qubit is shifted
relative to a neighboring ion by
$\delta\omega=|\kappa_1(B_{dc})-\kappa_0(B_{dc})|\frac{\mu_B}{\hbar}b\,\delta z$
where the distance between two ions is given by\cite{james}
$\delta z\approx z_0\frac{2}{N^{0.559}}$ and
$z_0=\left(\frac{e^2}{4\pi\varepsilon_0m\omega_z^2}\right)^{\frac{1}{3}}$,
and the coupling constants, $\kappa_1$ and $\kappa_0$
that characterize the particular hyperfine states chosen for the qubit
can be obtained from the Breit-Rabi formula.
To be concrete, we consider the $F=1$, $m_F=+1$ and $F=0$
hyperfine states\cite{huesmann} of $^{171}Yb^+$ in what follows.
In the weak field limit, $\frac{\mu_BB_{dc}}{E_{HFS}}\ll1$,
the Breit-Rabi formula gives $\kappa_1=1$ and $\kappa_0=0$, respectively,
whereas for $\frac{\mu_BB_{dc}}{E_{HFS}}=1$ we obtain
$\kappa_1=1$ and $\kappa_0=-0.89$ due to the non-linear Zeeman effect.

By choosing the magnetic field gradient $b$ appropriately,
the ions' qubit resonances can be well separated and any chosen ion
can be addressed by switching the frequency of the driving mw field.
If, in the usual addressing scheme (using focused laser beams), it were
possible to exclusively illuminate a single ion, that is, if  {\it resonant}
unwanted excitation could be avoided completely, then the remaining
source of unwanted excitation would be {\it nonresonant} excitation
of neighboring resonances (motional sidebands or carrier) 
of the ion being addressed. 
Our numerical studies show that
only the resonances next to the driven one contribute
appreciably to errors introduced by nonresonant excitation \cite{mintert}.
In the scheme proposed here, unwanted {\it resonant} excitation does not occur. 
We require the frequency separation between 
the sideband resonance
corresponding to the highest axial vibrational frequency $\omega_N$ \cite{radial}
of an arbitrary ion and the sideband resonance corresponding to $\omega_l$ (the bus-qubit)
of its neighboring ion to be larger than $\omega_l$. 
This corresponds 
to the frequency separation of resonances in the usual scheme and 
the probability  for spurious excitation of neighboring ions in the linear
chain is equal to or smaller than the probability of unwanted excitation 
of a resonance close to the desired one.
Given this requirement, the new scheme does not impose a new 
upper limit on the fidelity of basic quantum logic
operations due to an unwanted excitation, and an estimate
for the necessary B-field gradient is obtained from
$
b\ge\frac{\hbar}{2\mu_B}\frac{1}{|\kappa_1-\kappa_0|}
\left(\frac{4\pi\varepsilon_0m}{e^2}\right)^{\frac{1}{3}}
\omega_z^{\frac{5}{3}}
\left(4.7N^{0.56}+0.5N^{1.56}\right).
$
Here, $\omega_l=\omega_z$ and the highest vibrational frequency,
$\omega_N$ has been  approximated by the empirical law
$\omega_N=(2.7+0.5N)\omega_z$ valid for $5\le N\le100$ that was deduced
from numerical calculations of $\omega_N$ with $N$ ranging from 2 to 100\cite{mintert}.

In table 1 values of the field gradient necessary to spectrally separate
qubit resonances of $^{171}Yb^{+}$ are listed for different trap frequencies
and numbers of ions in one trap, respectively.
Magnetic field gradients of the magnitude required to separate the ions
resonances are well within capabilities of current technology.
Reichel, H\"ansel and H\"ansch\cite{reichel}, for example,
achieved gradients of about $300\frac{T}{m}$ over a distance of $50\mu$m
(which corresponds roughly to the axial extension of a string of 
$40$ $^{171}Yb^{+}$
ions at a trap frequency of $2\pi\times 500$kHz) using micro-fabricated conductors.
Gradients up to 8000 T/m are realistic in the near future \cite{folman}.

The  field gradient necessary to separate the ions resonances grows
with the number of ions stored in the trap.
This will limit the number of qubits available in a single trap \cite{usual}.
However, the scalability of a possible future ion trap quantum computer
does not rely on the storage of all qubits in a single trap.
Instead, arrays of traps communicating via 
``flying" qubits (photons)\cite{pellizzari} have been envisaged.
Communication between different traps can be established by the use
of photons that transfer quantum information, {\it e.g.}, via optical fibers.

We have investigated in detailed numerical calculations possible
detrimental effects associated with a magnetic field gradient 
applied to a linear ion trap\cite{mintert}. The dependence of the 
equilibrium position of each ion on magnetic forces that in turn 
depend on its internal state leads to a change of vibrational and 
internal transition frequencies when any one of the qubit 
internal states is changed. As a consequence, the transition 
frequency $\omega_0^k(\{a_j\})$, $j\in\{1,...,N\}$, $j\neq k$ of 
a given ion $k$ depends slightly on the internal states labeled 
$a_j$ of other ions. We calculated the mean transition frequency 
$\bar{\omega}_0^k=\frac{1}{2^{N-1}}\sum_{a_j,j\neq 
k}\omega_0^k(a_1,a_2,...)$ taking into account of the order of 
$N^2$ randomly chosen internal state configurations. The spread 
of $\omega_0$ around its mean value $\bar{\omega_0}$ is well 
characterized by a normal distribution with standard deviation 
$\sigma_k$ but which is cut off  at some value with typical size 
$2\sigma_k$ (maximum deviation from the mean value). The 
distribution of $\omega_0$ can be regarded as the width of the 
qubit transition. The  uncertainty in resonance frequency will 
only negligibly affect coherent manipulation of internal qubits 
and bus qubit as long as this uncertainty is much smaller than 
the Rabi frequency $\Omega_R$ between qubit states: A measure for 
the reliability of a quantum gate is the error $1-f$, with 
average fidelity $f=\frac{\Omega_R}{2\pi^2}\int_0^1d\alpha 
\int_0^{2\pi}d\varphi 
\int_0^{\frac{\pi}{\Omega_R}}dt|\langle\Psi_f|\Psi_r\rangle|^2$, 
where $|\Psi_r\rangle$ is the state obtained after an imperfect 
one-qubit rotation and $|\Psi_f\rangle$ denotes the final state 
that would be obtained if this operation were perfect. Averaging 
over initial states 
$|\Psi_i\rangle=\alpha|0\rangle+e^{i\varphi}\sqrt{1-\alpha^2}|1\rangle$ 
and pulse duration 
$1-f=\frac{41}{120}\frac{\sigma^2}{\Omega_R^2}$ is obtained 
($\sigma = 1/N\sum_{k=1}^{k=N} \sigma_k$) . The values of $1-f$ 
for $\Omega_R=\frac{1}{10}\omega_z$ listed in table \ref{table} 
show that the effect of the frequency change on the fidelity of 
quantum logic operations is well below  technological limits of 
current ion trap setups (for example \cite{sackett,naegerl2}).

All schemes devised for coherent manipulation of qubits
in usual traps can still be applied here.
In particular, fast quantum gates as suggested by Jonathan, Plenio and 
Knight\cite{jonathan} can be performed (the condition in our notation is 
$\Omega_R=\omega_l$). 
Sideband cooling to the vibrational ground state can be implemented
in the usual way, except that now microwave radiation is used
to drive the so-called red sideband of the hyperfine transition.
When, for example, Yb$^+$ is used, two commercial light sources
in conjunction with microwave radiation \cite{mintert} 
are sufficient for Doppler and sideband cooling of the bus-qubit, 
state preparation, coherent manipulation, 
and detection of qubits. 

In conclusion, the scheme proposed here permits coherent manipulation and
individual addressing of trapped ions using microwave radiation and can be
implemented using current ion trap technology in conjunction with
techniques from NMR spectroscopy.
Even multi-qubit operations should be possible using the present scheme.

We are indebted to H. Heyszenau for stimulating discussions and comments 
and to D. Reiss for careful reading of the manuscript. This work
was supported by the Deutsche Forschungsgemeinschaft. \\
$^*$Email: wunderlich@physnet.uni-hamburg.de.

\end{multicols}

\vspace{5em}
\begin{table}
\begin{center}
$
\begin{array}{|c|ccc|ccc|ccc|}
\hline
&
\multicolumn{3}{c|}{N = 10} &
\multicolumn{3}{c|}{N=20} &
\multicolumn{3}{c|}{N=40}   \\
& b\, \mbox{(T/m)}& \epsilon_c & 1-f &  b \, \mbox{(T/m)} & \epsilon_c & 1-f & 
b \, \mbox{(T/m)} & \epsilon_c & 1-f \\
\hline 
\omega_z/2\pi=100\mbox{ kHz} &
9.89 &  0.0075 & 3.4\! \cdot\! 10^{-6} &
22.1 & 0.012 & 5.2\! \cdot\! 10^{-5} &
54.7 & 0.021 & 1.1\! \cdot\! 10^{-3} \\
\omega_z/2\pi=1\mbox{ MHz} &
459 & 0.011 & 1.6\! \cdot \!10^{-5} &
1030 & 0.018 & 2.4\! \cdot\! 10^{-4} &
2540 & 0.031 & 4.9\! \cdot\! 10^{-3} \\
\hline
\end{array}
$
\end{center}
\caption{
The magnetic field gradient, $b$ ($b_0=0$) needed to separate the resonances of
$^{171}$Yb$^+$-ions, the coupling constant $\epsilon_c$ 
(analogous to the Lamb-Dicke-Parameter), and
the average error, $1-f$ for an arbitrary qubit rotation 
(for Rabi frequency $\Omega_R=\frac{1}{10}\omega_z$)
for different trap frequencies and numbers of ions. Gradients up to 
8000 T/m are within reach of current experiments \protect\cite{reichel,folman}.}
\label{table}
\end{table}


\begin{references}
\bibitem{radial} Since 
there is no gradient in the radial direction, microwave
radiation does not couple to radial modes.
Therefore only axial modes have to be considered.
\bibitem{usual} In usual traps, too, the number of qubits is limited, among other
restraints, by the distance between them that scales as 
$\omega_z^{-2/3}N^{-0.56}$ \cite{james}.
\end{references}
\end{document}